# Tunable cryogenic THz cavity for strong light-matter coupling in complex materials

Giacomo Jarc[1,2], Shahla Yasmin Mathengattil[1,2], Francesca Giusti[1,2],
Maurizio Barnaba[2], Abhishek Singh[3], Angela Montanaro[1,2], Filippo Glerean[1,2],
Enrico Maria Rigoni[1,2], Simone Dal Zilio[4], Stephan Winnerl[3], and Daniele Fausti[1,2]

[1]Department oh Physics, Università degli Studi di Trieste, 34127 Trieste, Italy

[2]Elettra Sincrotrone Trieste S.C.p.A., 34127 Basovizza Trieste, Italy

[3]Institute of Ion Beam Physics and Materials Research, Helmholtz-Zentrum Dresden-Rossendorf, Bautzner Landstrasse
400, 01328, Dresden, Germany

[4]CNR-IOM TASC Laboratory, Trieste 34139, Italy

We report here the realization and commissioning of an experiment dedicated to the study of the optical properties of light-matter hybrids constituted of crystalline samples embedded in an optical cavity. The experimental assembly developed offers the unique opportunity to study the weak and strong coupling regime between a tunable optical cavity in cryogenic environment and low energy degrees of freedom such as phonons, magnons or charge fluctuations. We describe here the setup developed which allows the positioning of crystalline samples in an optical cavity of different quality factor, the tuning of the cavity length at cryogenic temperatures and its optical characterization with a broadband time domain THz spectrometer (0.2 – 6 THz). We demonstrate the versatility of the setup by studying the vibrational strong coupling in $CuGeO_3$ single crystal at cryogenic temperatures.

## I. INTRODUCTION

The possibility of exploiting light-matter interaction to control and manipulate material properties has generated much interest in the last decades. In the 1940s Purcell discovered that the emission of a given emitter could be modified when it is placed in a resonant optical cavity [1]. This was the first demonstration of the more general principle that the properties of an emitter can be control by engineering its electromagnetic environment. In condition where the lifetime of the photon in the cavity is short a *weak coupling* regime can be obtained and results in the experimental observation of either an enhancement [2] or suppression [3,4] of the rates of spontaneous emission. If the optical confinement is sufficiently strong, the presence of the cavity surrounding the emitter can be described as a coherent evolution with a photon repeatedly absorbed and reemitted by the emitter. In this limit, called *strong coupling* limit, the strength of the coupling between the resonator is determined mainly by two factors: the oscillator strength of the optical transition which is fixed by the nature of the emitter and the quality factor of the cavity which can be tuned and determines the photon lifetime inside the cavity. In the strong coupling limit, the light-matter interaction overcomes the dissipative processes occurring in the uncoupled systems and the wavefunctions of the material excitations and the photon inside the cavity are coherently mixed, forming hybrid light-matter states called polaritons [5,6]. This results in the splitting of the material targeted excitation into two bands dubbed upper polariton (UP) and lower polariton (LP) carrying the spectral weight of all the emitters inside the cavity.

The core aspect that has raised much interest in light-matter hybrids in the strong coupling regime is the fact that polaritonic states can be regarded as highly delocalized states. This delocalization arises from the fact that many emitters can be placed in the optical mode volume, and therefore can simultaneously interact with a common cavity mode which induces quantum correlations among the spatially separated emitters. For this reason, while the hybridization between light and matter was originally realized and comprehended in isolated atomic and molecular systems [1-4, 7-9], it has more recently been extensively studied in inorganic and organic semiconductors excitons [10-12], phonons [13–20], magnons [21]. Exploiting the strong coupling limit is emerging as a promising tool to control material functionalities in different physical-chemistry settings. It has been shown that vibrational strong coupling can affect chemical reactivity [22,23], conductivity [24], molecular structure [25] and charge and energy transfer [26–29].

The purpose of this work is to report the realization of an experimental setup to extend the studies of cavity-mediated light-matter coupling to crystalline solids.

We present here a versatile setup suitable for the study of light-matter interaction in the THz range in cavity confined systems at low temperatures. The unique feature of the setup consists in the possibility of tuning the cavity resonance in a cryogenic environment thus enabling to study the coupling at low temperature of a vast number of low energy excitations (phonons, magnons, charge fluctuation and many others). This has been realized in a cryogenic chamber in which the cavity mirror moving system is thermally decoupled with respect to the cryogenic cavity environment. We combined the cryogenic cavity assembly with a broadband time domain THz spectrometer which allows at the same time the characterization of the cavity optical characteristics (fundamental mode and quality factor) and the study of the linear response of the light-matter hybrids at low temperatures. We commissioned the setup and used it to study the hybridization at 80 K of an IR-active THz phonon in $CuGeO_3$. We proved that, thanks to the high oscillator strength of the vibrational excitation, a strong coupling regime can be reached even with a cavity with low quality factor ($Q \sim 6.8$).

## II. SETUP DESIGN AND CHARACTERIZATION

### A. Cryogenic cavity assembly

A detailed scheme of the built variable-length cryogenic THz cavity is presented in Figure 1. This is composed of two cryo-cooled piezo-controlled movable mirrors between which the sample is inserted. The movement of each of the two cavity mirrors is ensured by three piezo actuators (N472-11V, Physik Instrumente) with a total travel range of 7 mm and a minimum incremental motion of 50 nm with a designed resolution of 5 nm. The independent movement of each of the three piezo actuators ensures the independent horizontal and vertical alignment of the mirrors while the simultaneous motion of the three results in a rigid translation of the whole mirror. Importantly, since both the mirror positions are controlled by the piezoelectric mechanics, the setup includes both the possibility of tuning independently the cavity length and the sample position with respect to the mirrors. The tunability of the cavity length sets the frequency of the cavity fundamental mode. Instead, the tunability of the sample position with respect to the mirrors allows to maximize the coupling of the cavity photons with the targeted excitation, since the coupling energy scales with the absolute cavity field [5]. The mirrors are mounted on copper holders and they are cryo-cooled by mean of copper braids directly connected to the cold finger of the cryostat. Since the piezo actuators temperature operational range is 283-313 K, the piezo actuators are thermally decoupled from the mirror supports. The thermal decoupling is realized by placing between the piezo actuators and the mirror holders a PEEK disc on which the actuators actually act and three ceramics cylinders. These materials are thermal insulators and they have a low thermal expansion coefficient in the operational temperature range of the cryostat (10 K – 300 K). These features ensure the mirrors to be thermally insulated as well as an alignment stability of the cavity in the operational temperature range. We tested the setup in nitrogen-cooled conditions and proved that in the temperature range 80 K – 300 K the thermal decoupling between the cryo-cooled mirrors and the piezo actuators is efficient, thus making the setup suitable to perform cavity length-dependent studies in cryogenic environment.

The cavity semi-reflecting mirrors were fabricated by evaporating a thin bilayer of titanium: gold (2-10 nm) on a 2 mm thick crystalline quartz substrate, resulting in a transmission amplitude of 20 % across the THz spectral range of the experiment with no apparent spectral features. In detail, the deposition of the thin film coating has been achieved by classical E-beam evaporation: the substrates were firstly cleaned by standard procedure based on RCA-1 ($NH_4OH$-$H_2O_2$-$H_2O$ 1:1:5, 75°C, 10'), rinsing by DI water and dry under $N_2$ blow. Right before the transfer in the evaporator chamber, the substrates were treated by oxygen plasma (P:20W, B:50V, t:1'). The first 2 nm thin layer of Ti was used to increase the adhesion of the following Au layer, achieved with a deposition rate of 0.4 Å/sec.

The sample is mounted between the mirrors in a copper sample holder directed connected to the cold finger of the cryostat and sealed between two silicon nitride membranes (LP-CVD grown) with a window size of 11 mm x 11 mm and a thickness of 2 µm (Figure 1B). The membranes are supported on a 13 mm x 13 mm silicon frame which has a thickness of 500 µm. Importantly, the membranes are transparent in the THz frequency range employed in the experiments and does not show any spectral dependence.

The chamber, depicted in Figure 1A, is mounted on a flow cryostat which is supported by a mechanical assembly allowing for the movement of the whole sample in the $x$, $y$ and $z$ directions. We stress that the mechanical translation of the sample is particularly crucial for the experiment since it allows to perform THz transmission measurements of the empty cavity by simply moving the vertical/horizontal position of the whole chamber of Figure 1A. The chamber shown in Figure 1A is enclosed in a vacuum chamber allowing optical access for transmission. The cryostat windows are two 2 mm crystalline quartz windows, which are suitable for the THz range. The vacuum conditions are ensured via a turbo pumping system (Pfeiffer Hi-Cube). Pressures of $10^{-6}$ mbar can be reached at room temperature, while at cryogenic temperature the typical working pressure is $10^{-7}$ mbar. The temperature is read on the sample holder by mean of a cryogenic silicon diode. A temperature controller provided with a feedback circuit enables to modify the sample temperature so that a complete temperature scan can be performed at a fixed cavity length.

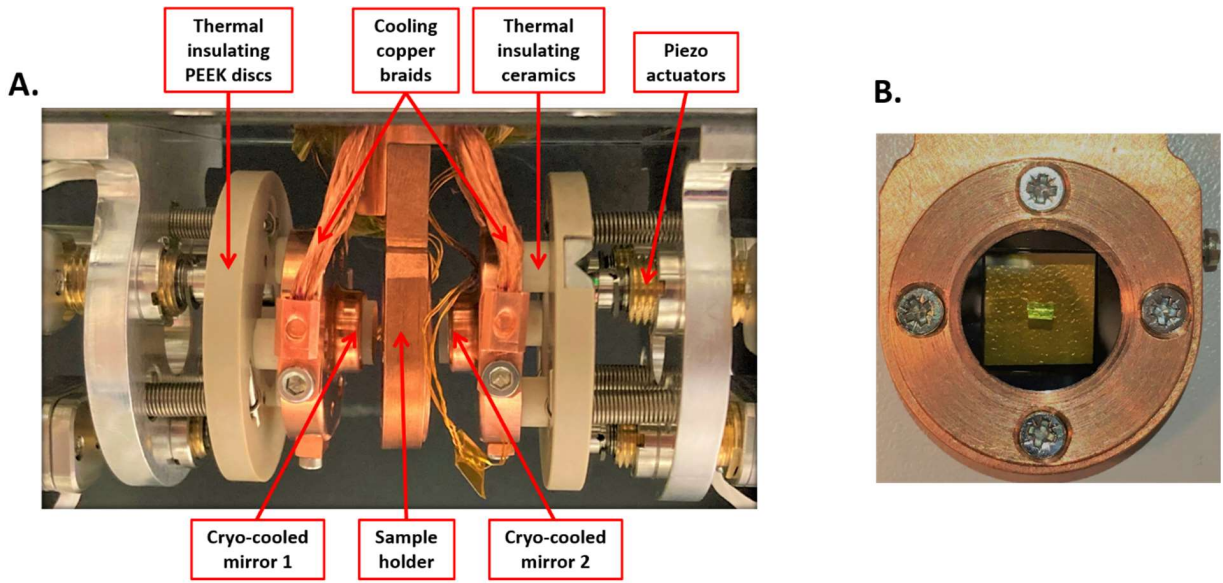

FIG. 1. A. Cryogenic cavity detailed illustration. The cavity mirrors are kept in thermal contact with the sample holder while they are insulated from the piezoelectric mechanics B. The sample is mounted between two transparent silicon nitride membranes of 2 μm thickness.

## B. Optical setup

The layout of the THz spectrometer is shown in the schematic diagram in Figure 2A. Ultrashort laser pulses (50 fs pulse duration, 745 nm central wavelength) from a commercial 50 KHz pulsed laser + optical parameter amplifier (OPA) system (Pharos + Orpheus-F, Ligth Conversion) are split in two to form an intense optical beam for THz generation (6 μJ/pulse) and a weak readout pulse (< 100 nJ/pulse) for time-resolved electro-optical sampling.

Single-cycle THz pulses are generated via the acceleration of the photoinduced carriers in a large-area GaAs-based photoconductive antenna (PCA). The THz emitter is fabricated on a semi-insulating GaAs substrate by depositing the metal electrodes having an interdigitated finger-like structure. Each electrode is ~ 1 cm long and 10 μm wide. The gap between two nearby electrodes, which is also the active region, is 10 μm. The total area of the emitter is 1 cm x 1 cm. Details about the emitter design and fabrication can be found in [30–33]. Due to such a narrow electrode gap, a bias of just a few volts on the electrodes creates an electric field of the order of a few kV/cm in the active region. Now, photoexcitation of active regions creates charge carriers in the GaAs, which accelerate due to the presence of an applied electric field and emit THz radiation. Polarization of the emitted THz is parallel to the applied electric field. To avoid the destructive interference of THz radiated from two neighbouring active regions, each alternate active region is covered with a metallic layer to avoid the photoexcitation and hence out-of-phase THz generation from those regions.

The acceleration of the free carriers induced by the pump is achieved by biasing the PCA with a square-wave bias voltage $V_{BIAS}$ triggered with the laser at a frequency of 1.25 KHz. We employed a biasing square wave with a voltage peak of 8.0 V and a 50 % duty cycle. For an efficient THz generation using 6 µJ pump pulse energy, an area of around 6 mm diameter on the 1 cm$^2$ large emitter is illuminated using a collinear pump beam. Since the diameter of the excitation area is comparatively much larger than the THz wavelength, the radiated THz beam has a similar wavefront as the pump beam on the emitter and hence follows the same beam path as the pump beam.

The emitted collimated THz beam is then focused on the sample mounted inside the cavity which is placed in the focal plane of two off-axis parabolic mirrors (OPM). The THz field and the readout pulse are then combined and focused on a 0.5 mm ZnTe crystal which acts as electro-optical crystal. After the electro-optical crystal the probe beam, variable delayed in time through a translation stage (TS), is analyzed for its differential polarization changes induced by the THz in the ZnTe crystal which maps the time evolution of the ultrafast THz field. This is done by standard Electro-Optical Sampling (EOS) [34,35], i.e. by splitting the two probe polarizations with a Wollaston Prism and measuring the differential intensity recorded on a pair of photodiodes. The resulting differential signal is then detected using a lock-in amplifier (SR830, Stanford Research System) referenced at the frequency of the bias voltage ($V_{BIAS}$). We estimated the signal-to-noise ratio of the detected THz field to be 4.6 x 10$^4$.

The entire system is purged with nitrogen to eliminate THz absorption coming from the water vapor in the ambient atmosphere. We show in Figure 2B the measured electric field of the generated THz pulse and its calculated Fourier spectrum (Figure 2C). As can be seen, the input field is indeed a nearly single-cycle THz pulse with spectral content reaching 6 THz, as highlighted in the logarithmic scale plot in the inset of Figure 2C.

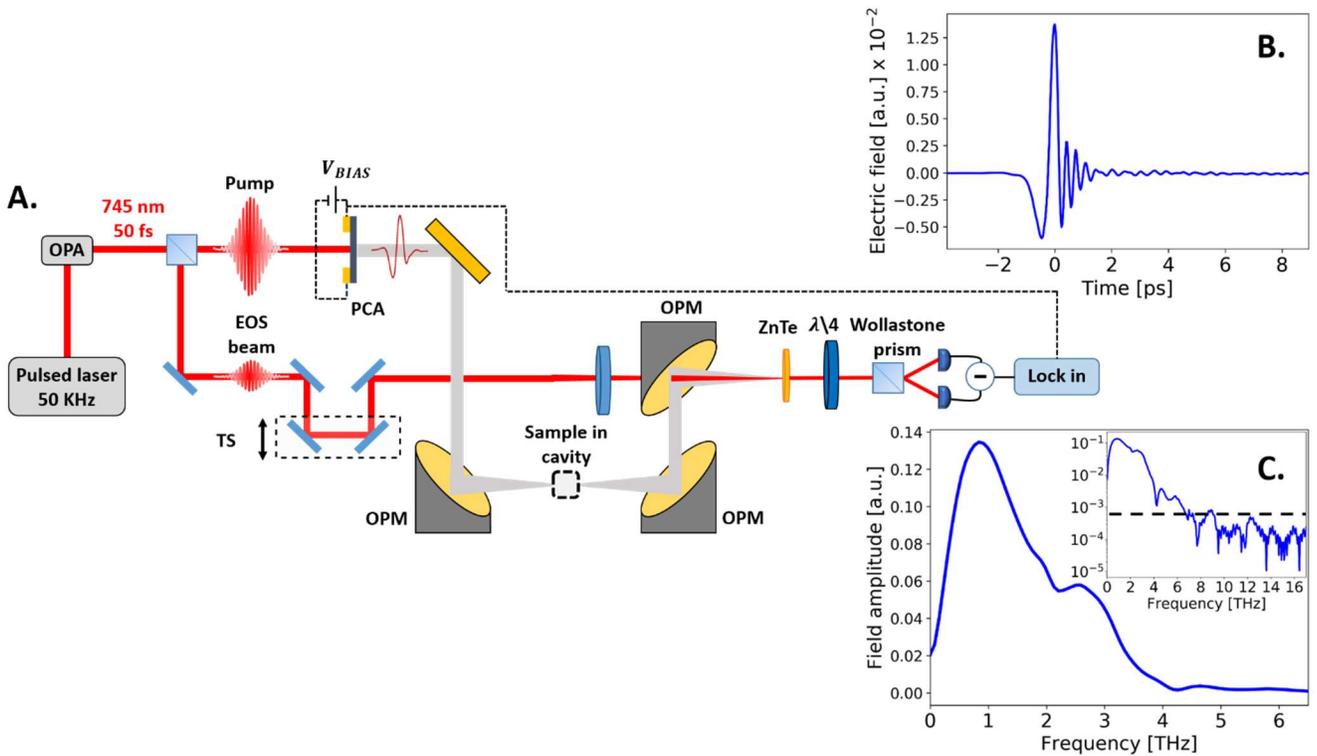

FIG. 2. A. Sketch of the built THz time domain spectrometer. B. Free space nearly single-cycle THz field employed in the experiments detected trough Electro-Optical Sampling (EOS) in a 0.5 mm ZnTe crystal. C. Fourier transform of the nearly single-cycle THz field in free space. In the inset the Fourier spectrum is plotted in logarithmic scale to highlight the spectral content of the THz field up to approximatively 6 THz. The black dashed line in the logarithmic plot indicates the noise level.

## C. Empty cavity characterization

In this section we present the characterization of the response of the empty cavity at 80 K, i.e. when the THz field passes only through the silicon nitride membranes within the mirrors. With this characterization the quality factor of the cavity can be determined. The cavity quality factor is a crucial parameter for the experiment setting the photon lifetime inside the cavity and hence the coupling strength between the cavity mode and the targeted material resonance. In order to minimize the photon losses and hence maximizing the cavity quality factor, the two cavity mirrors was set parallel each other and perpendicular to the THz incoming beam. This was obtained by aligning the multiple reflections of the pump beam, which is made collinear with the THz by the generation process. The alignment was then finely tuned by maximizing the THz field peaks in the time domain trace associated to the multiple reflections of the THz beam within the cavity.

The results of the characterization are presented in Fig. 3 where we plot the time domain THz field transmitted through the Fabry-Perot empty cavity and the corresponding spectral content for three representative values of the cavity length.

The transmission spectra are obtained by taking the ratio between the Fourier spectrum of the time domain THz traces shown in Fig. 3A and the reference free space spectrum shown in Fig. 2C. The time-dependent detected fields (Fig. 3A) show that when the nearly single-cycle THz pulse passes through the cavity is repeatedly reflected by the mirrors with a round-trip time set by the cavity length. This causes the nearly single-cycle THz field to be stretched to a multi-cycle decaying oscillating field with a decay time set by the cavity quality factor. This results in transmission spectra (Fig. 3B) exhibiting interference Fabry-Perot modes with their frequency determined by the equation $\nu_m = \frac{c}{2nL}m$, where L is the length of the cavity, n the refractive index of the medium inside the cavity and m the mode number. The estimated quality factor of the cavity at 80 K, defined as the ratio between the fundamental cavity mode and its bandwidth at a fixed cavity length, is Q = 6.8. We have employed the cavity configuration with L = 103 μm for this estimation, which has its fundamental mode at $\nu_1 = 1.45$ THz, since this is the frequency of the targeted phonon excitation discussed in the manuscript (Section III).

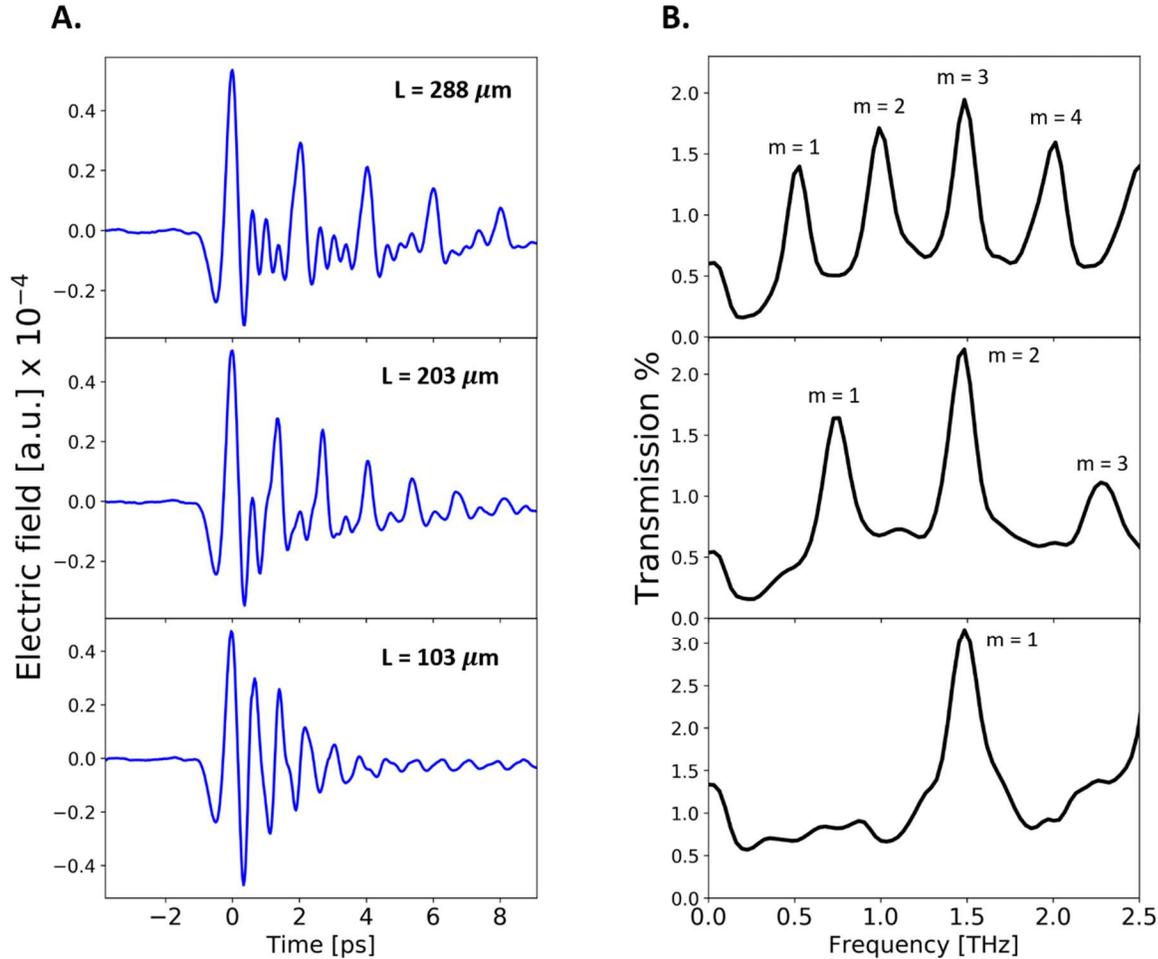

FIG. 3 THz spectroscopy of the empty cavity at 80 K for three representative distances L between the cryo-cooled mirrors A. Measured time-resolved THz fields transmitted through the empty cavity. B. Corresponding transmission spectra, showing the tunability of the cavity fundamental resonance mode (m = 1).

## III. MEASUREMENTS

We exploited the setup to demonstrate vibrational strong coupling in $CuGeO_3$ in cryogenic environment at 80 K. $CuGeO_3$ is an insulating crystal belonging to the family of cuprates. Its room temperature crystalline structure is depicted in Figure 4A and takes the name of "normal" or "undistorted" phase. The building blocks of the crystal structure are $Cu^{2+}$ and $Ge^{4+}$ chains parallel to the $c$ axis. These chains are linked together through the O atoms and form layers parallel to the $b$-$c$ plane weakly coupled along the $a$ axis [36–38]. $CuGeO_3$ is highly studied for its peculiar magnetic behavior revealing the onset of a spin-Peierls phase (below 14 K) in which the lattice distortion in accompanied by the formation of a spin-singlet ground state and the creation of an energy gap in the spectrum of magnetic excitations [38–40].

A full review of the physics of $CuGeO_3$ is beyond the scope of this manuscript. $CuGeO_3$ is chosen to test the potentiality of the developed setup because it has a strong Cu-O IR-active vibrational mode in the THz range which shows a monotonic red-shift in the normal phase from 300 K to 14 K [38]. This can be measured only with a THz electric field polarized perpendicular to the magnetic chains (and hence lying on the $b$ axis) while no absorption is present at this photon energy when the THz radiation is polarized parallel to the chains. For this reason, the sample was oriented in order to have the $b$ axis lying on the same direction of the THz electric field.

The frequency-dependent transmission at 80 K of the 20 μm thick $CuGeO_3$ crystal employed in this study is presented in Figure 4B. The latter was measured in the open cavity configuration, i.e. when the distance between the two mirrors is such that the fundamental cavity frequency lies far below with respect to the phonon frequency and the resulting transmission can be regarded as the free-space one with only a damping coefficient due to the semi-reflecting mirror absorption. The 80 K $b$ axis transmission spectrum presented in Figure 3B shows a sharp absorption at 1.45 THz with a free-space linewidth of $\gamma_{pho}$ = 76 GHz full-width half-maximum.

After having characterized the bare phonon response, we examined the 80 K response of the sample placed in the center of a cavity whose fundamental frequency mode is resonant to the vibrational mode at 1.45 THz. As highlighted in Section I, placing the sample in the middle of the cavity is crucial since we expect the phonon coupling with the cavity fundamental mode to be maximum at the maximum of the cavity field, which is the cavity center for the ground state mode. The sample was centered in the middle of the cavity exploiting the THz time domain trace in the open cavity configuration. This was achieved by temporally overlapping the THz field peak associated to the reflection between the sample and the first mirror and the one associated to the reflection between the sample and the second mirror. The transmission spectrum of the cavity resonant to the CuGeO$_3$ phonon mode is presented in Figure 4C (blue line) together with the free space phonon transmission (red dashed line) and the empty cavity transmission at the phonon frequency (black dashed line).

The hybrid sample-cavity system exhibits a splitting in its spectral response around the phonon frequency with a frequency separation greater than both the dissipative response of the free-space phonon and the cavity linewidth, which quantifies the photon dissipative rates inside the bare cavity. This indicates that a strong coupling regime [4–21] can be reached and the two bright peaks can be associated to vibro-polariton states. The lower (upper) energy peak in transmission is dubbed lower (upper) polariton (LP, UP). The energy separation between the two vibro-polaritonic states, dubbed Rabi splitting, is estimated to be $\Omega_R = 0.32$ THz. The measured Rabi splitting is approximatively the 22 % of the bare phonon frequency, placing the hybridized system close to the ultrastrong coupling regime [6,15].

The fingerprints of the strong coupling regime are visible also in the time-dependent THz field exiting the cavity. To illustrate this we present in Figure 4D the time domain THz field exiting the phonon-resonant cavity filtered in the range 0.5-2.3 THz. Note that we applied this Fourier filter [15,21] in order to exclude from the resonant emitted field the response of the second-order uncoupled cavity mode centered at 3.0 THz which lays inside the bandwidth of the input THz pulse and is therefore superimposed in the time domain trace. This filtering procedure of the time domain data allows to examine the evolution of the emitted field at frequencies around the resonance and hence highlight more clearly the coherent energy exchange between photonic and phononic degrees of freedom associated to strong coupling [5–22]. At resonance, the signal exiting the cavity is an exponentially decaying field modulated by a periodic beating with a period $1/\Omega_R = 3.1$ ps. This periodic modulation corresponds to coherent Rabi oscillations in the cavity and indicates that there is a coherent energy exchange between phonons and photons at a rate $\Omega_R = 0.32$ THz occurring inside the cavity. We note indeed that if there were no splitting associated to the strong coupling regime, the cavity and the vibrational mode would be frequency-degenerate at resonance and hence exhibit no temporal beating.

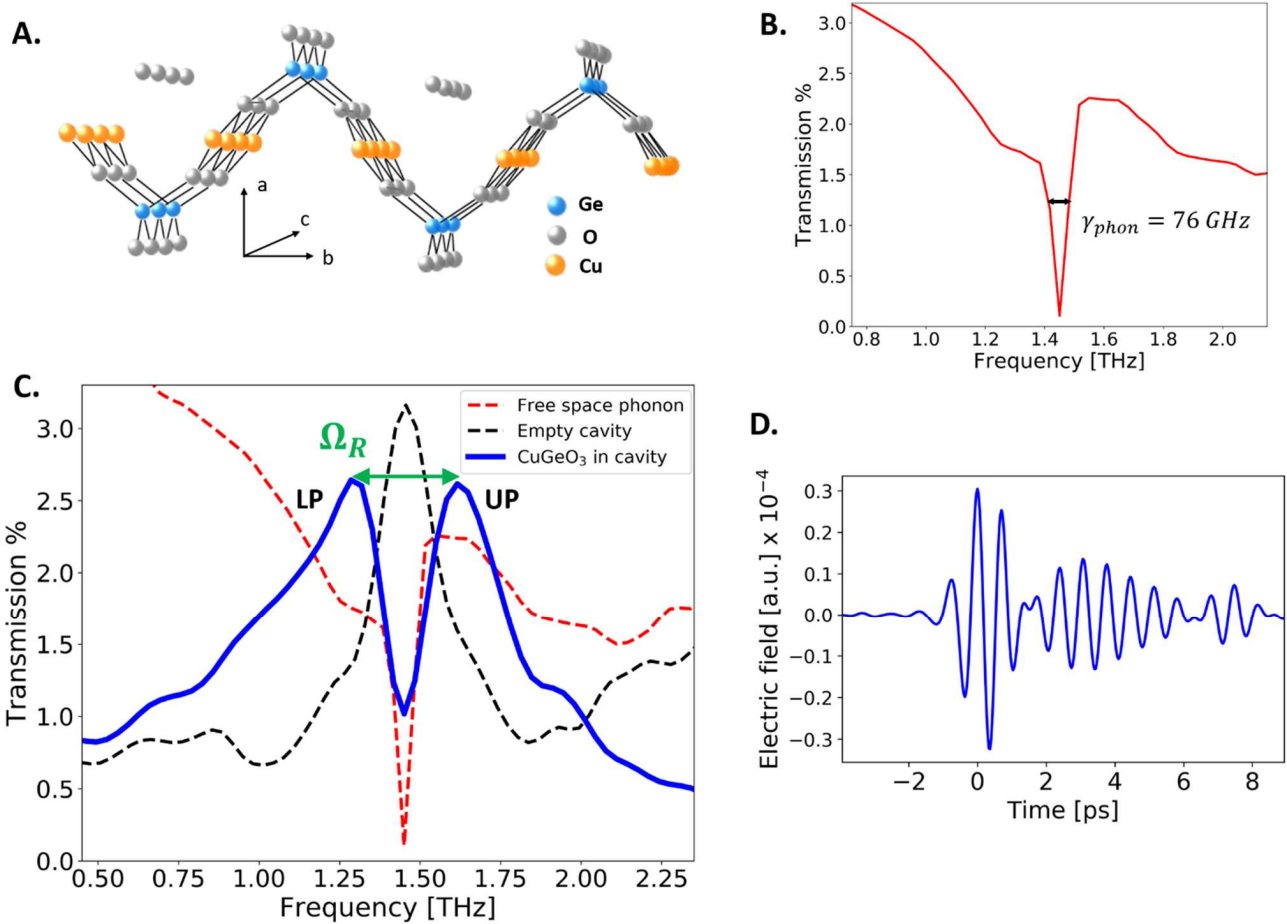

FIG. 4. Demonstration of the phonon strong coupling regime in CuGeO₃ at 80 K. A. CuGeO₃ crystal structure in the normal phase. B. Open cavity CuGeO₃ transmission at 80 K of THz light polarized along the *b* axis, showing an IR-active mode at 1.45 THz with a linewidth $\gamma_{phon}$ = 76 GHz. C. Static transmission of CuGeO₃ at 80 K in a cavity of length with fundamental frequency resonant to the phonon one. Strong coupling between the phonon and the cavity mode results in a spectral splitting in two new modes: Upper Polariton (UP) and Lower Polariton (LP) with a separation $\Omega_R$ greater than either the free space phonon linewidth (red dashed curve) and the empty cavity linewidth at the resonance frequency (black dashed curve). D. Time-dependent THz field at resonance filtered in the range 0.5-2.3 THz showing coherent Rabi oscillations associated to the strong coupling regime.

In order to examine the tuneability of the resonance of the cavity we measured the anti-crossing between the two polaritonic states. This is a distinctive feature of strong coupling and corresponds to the creation of two separate polaritonic branches that do not intersect when the cavity resonance lies within the absorption band of the targeted excitation [5–22]. We tracked the emergence of the two polariton branches by symmetrically varying the position of the two mirrors around the cavity mode which is therefore tuned across the phonon frequency. In Figure 5A we plot the THz transmission for each mirror configuration and the obtained dispersion of the polaritonic branches as a function of the cavity fundamental frequency. Figure 5B reports representative transmission spectra for different detunings around the phonon frequency, which correspond to vertical cuts of Figure 5A. The measured evolution of the transmission spectra shows that when the cavity is detuned away from the vibrational absorption frequency, the frequencies of the polaritonic modes shift with respect to the resonant case, and their relative spectral weight is also modified. Indeed, as highlighted in Figure 5A, when the energy of the cavity fundamental mode is different with respect to the phonon one, the energies of the two polariton branches approach the ones of the uncoupled system (red and black dashed lines in Figure 5A), while in resonant condition the difference between the polariton energies and the uncoupled systems ones is maximum. This results in the avoided crossing around the phonon frequency highlighted in Figure 5A. In Figure 5C we present the evolution of the filtered time domain THz fields exiting the cavity for different detunings $\Delta\omega$ around the phonon frequency. We show that tuning the cavity mode away from the phonon resonance is mapped in the time domain with a damping of the coherent Rabi oscillations with respect to the resonant case $(\Delta\omega = 0)$.

Figure 5D reports the comparison of the resonant cavity-phonon response at 80 K and 295 K. A comparison between the transmission spectra reveals a red-shift of the polariton splitting of about 0.04 THz between the 295 K and the 80 K hybrid system response. This frequency shift is consistent with the changes of the phonon frequency with temperature in the bulk $CuGeO_3$ in normal phase [37].

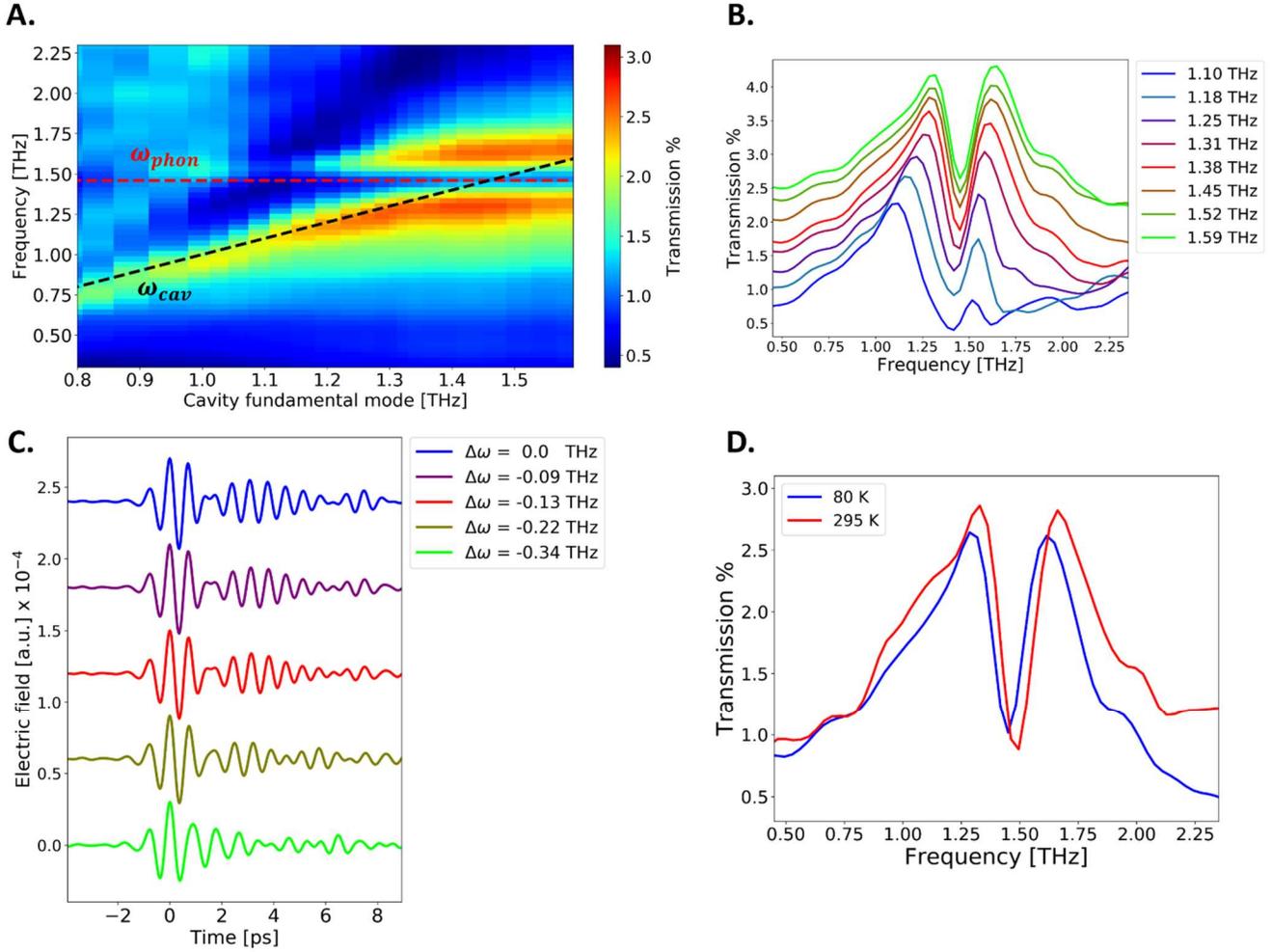

FIG. 5. A. Phonon-polariton dispersion in the strong coupling regime of CuGeO$_3$ at 80 K. The red dashed line marks the uncoupled phonon frequency $\omega_{phon}$ = 1.45 THz while the black dashed one the uncoupled cavity mode $\omega_{cav}$. B. Evolution of the cavity transmission spectra at 80 K with cavity fundamental frequencies (indicated in legend) across the phonon resonance C. Evolution of the filtered THz fields exiting the cavity at 80 K for different cavity detunings $\Delta\omega = \omega_{cav} - \omega_{phon}$ D. Comparison of the strongly coupled cavity-phonon transmission at 80 K and 295 K. The shift of the polariton splitting frequency is consistent with a phonon frequency shift.

## CONCLUSIONS

In this manuscript we have reported the development and commissioning of a novel setup to study the terahertz optical properties of low energy degrees of freedom in solid samples coupled with a tunable optical cavity in cryogenic environment. The unique feature implemented in the setup lies in its capability of tuning the cavity resonance at cryogenic temperatures. This is crucial to target the light-matter coupling of different material excitations and study how their coupling with an optical cavity mode may affect the material macroscopic properties. The light-matter hybrids are characterized with a broadband THz nearly single-cycle field generated in a photoconductive antenna and with a frequency content up to 6 THz. We tested the performance of the setup by characterizing the hybridization at 80 K of an IR-active phonon of CuGeO$_3$. We showed that, thanks to the high oscillator strength of the phonon mode, vibrational strong coupling can be reached in the system even with low quality factor cavities (Q = 6 8). The measured Rabi splitting obtained at 80 K is estimated to be approximatively the 22 % of the bare vibrational frequency. We lastly varied the temperature of the system and detected a red-shift of the polariton anti-crossing frequency going from higher to lower temperatures, consistent with a shift of the bare phonon resonance. The capability of the setup of tuning the cavity resonance combined with its capability of performing temperature-dependent studies in a wide range of cryogenic temperatures makes it a versatile platform for the study of how light-matter hybridization of different low energy excitations may affect the macroscopic properties of complex materials. The use of broadband THz fields is particularly crucial in this sense as allows the simultaneous characterization of the empty cavity and the cavity hybridized with the contained material. The described setup enables the study of the light-matter hybridization with complex solid materials where a plethora of low energy cavity-induced effects have been predicted: as superconductivity [41,42], magnetism [43,44], and charge transport [28,45].

**ACKNOWLEDGMENTS**

**DATA AVAILABILITY**

The data that support the findings of this study are available from the corresponding author upon reasonable request.